\documentclass{article}
\usepackage[preprint]{spconf}
\usepackage{amsmath,graphicx,url}
\usepackage[hidelinks]{hyperref}
\usepackage{fancyhdr}
\usepackage{booktabs}
\usepackage{multirow}

\title{LEARNING TO JOINTLY TRANSCRIBE AND SUBTITLE FOR END-TO-END SPONTANEOUS SPEECH RECOGNITION}
\name{Jakob Poncelet, Hugo Van hamme}
\address{KU Leuven \\ 
     Department Electrical Engineering ESAT-PSI\\
     Kasteelpark Arenberg 10, Bus 2441, B-3001 Leuven, Belgium\\
     }

\begin{document}

\thispagestyle{fancy}

\maketitle

\begin{abstract}
TV subtitles are a rich source of transcriptions of many types of speech, ranging from read speech in news reports to conversational and spontaneous speech in talk shows and soaps. However, subtitles are not verbatim (i.e. exact) transcriptions of speech, so they cannot be used directly to improve an Automatic Speech Recognition (ASR) model. We propose a multitask dual-decoder Transformer model that jointly performs ASR and automatic subtitling. The ASR decoder (possibly pre-trained) predicts the verbatim output and the subtitle decoder generates a subtitle, while sharing the encoder. The two decoders can be independent or connected. The model is trained to perform both tasks jointly, and is able to effectively use subtitle data. We show improvements on regular ASR and on spontaneous and conversational ASR by incorporating the additional subtitle decoder. The method does not require preprocessing (aligning, filtering, pseudo-labeling, ...) of the subtitles.

\end{abstract}

\begin{keywords}
speech recognition, multitask learning, end-to-end, subtitles, spontaneous speech
\end{keywords}

\section{Introduction}
\label{sec:intro}
Speech recognition aims to identify the words in a spoken audio fragment and transcribe what has been said into written text. Traditionally, Automatic Speech Recognition (ASR) systems made use of a limited vocabulary of words and pronunciations, and consisted of several components to model speech statistically. More recently, data-driven end-to-end ASR systems have been able to directly convert a sequence of audio features into a sequence of (sub)words or characters.

ASR requires a large dataset with careful transcriptions of spoken utterances. While TV subtitles seem the obvious candidate, as they are so abundant and manually annotated, they are not \textit{verbatim} (i.e. exact) transcriptions of what was said. Subtitles are optimised for readability on screen, and are therefore often shorter and cleaner (e.g. no repetitions, no hesitations). Consequently, the timings of subtitles on screen can be inaccurate compared to the spoken sentences. On top of that, subtitles are ``clean", with edits 
to correct non-grammatical 
sentences and poor or dialectal word choice. 

Nevertheless, being able to use TV subtitles would have many advantages. Subtitles can cover a very broad domain, from read speech in broadcast news to spontaneous and conversational speech in talk shows, interviews, soaps, sitcoms, etc. This is in contrast to the labeled speech datasets which often contain prepared speech (e.g. audiobooks in LibriSpeech \cite{librispeech}). Subtitles cover many languages, and they have been manually annotated by human annotators. Additionally, accented speech and strong dialects are more common among free speech on TV, and less covered in current speech datasets.

Reducing the need for well-labeled speech data is a long-standing problem. The performance of ASR has seen tremendous improvements from pre-training on unlabeled data in a self-supervised manner \cite{wav2vec2, hubert, wavlm}. Unlabeled speech data is plentifully available, and can even be (pseudo-)labeled by pre-trained models to iteratively improve models by self-training \cite{selftraining}. Furthermore, pre-training reduces the need for large amounts of labeled data and has significantly advanced performance in low resourced languages
, on account of cross- and multilingual positive transfer \cite{XLSR} from learned high-level acoustic feature representations to the target language. 

In this work, we propose a model that jointly performs automatic (verbatim) speech recognition and automatic subtitling. Besides solving two useful tasks at once, the model is able to make use of the subtitle data. We apply a multitask learning (MTL) framework \cite{MTL1, MTL2} with a shared encoder and two decoders, inspired by end-to-end speech translation where machine translation and ASR are additional subtasks with independent \cite{tied2018} or interconnected decoders \cite{Le2020, interactive2020} for each task. We show that the introduction of the subtitle decoder and subtitle data improves the performance of the end-to-end ASR model on both a regular ASR benchmark and on a spontaneous speech dataset, without construction of a spontaneous ASR dataset. Experiments are performed on Belgian Dutch (Flemish), a medium-resourced language with strong regional dialects. Finally, we note that parallel work on pre-training with unlabeled data for learning representative features (e.g. for Belgian Dutch \cite{poncelet}), is 
complementary to this work, and these pre-trained feature representations can be plugged into the front-end of an ASR system. 

\section{Related work}
Most research on the use of subtitle data can be divided into two strands. On the one hand, subtitles (either detected on screen in video's \cite{Bang2020} or gathered separately) are refined and strongly filtered to build a corpus for ASR \cite{Lamel, Bang2017}, e.g. based on matching alignment \cite{CU_MGB1, Shintaro_ICASSP21, LIUM_MGB2} or thresholding on posterior probability \cite{SU_MGB1}. 
On the other hand, the output of ASR systems can be adapted with models into readable subtitles, by compression \cite{iwslt1} or with a (genre-dependent) language model \cite{CRIM_MGB1}, or directly translated into subtitles in a different language \cite{icalt1}. The Multi-Genre Broadcast (MGB) challenge inspired some of these works, with the MGB-1 challenge \cite{MGB1} providing English subtitles and the MGB-2 challenge \cite{MGB2} Arabic subtitles, and related competitions, e.g. the IberSpeech-RTVE challenge on Spanish Broadcast media \cite{IberSpeechRTVE}. Finally, there has also been research on using weak supervision from contextual text information in social media video's to improve ASR \cite{WeaklyICASSP2020}, however these annotations are further away from verbatim transcriptions than subtitles.

The work in this paper distinguishes itself by concurrently solving both the ASR and subtitling task, and, without having to adapt or refine the data (which can be done complementary to our method in a pre-processing stage, e.g. subtitle alignment \cite{lightsup_align}), improving the current ASR system in all domains. No parallel data is used, i.e. data for which there are both verbatim and subtitle transcriptions, as it is not generally available in most languages.



\section{Method}
The proposed model adapts the end-to-end hybrid CTC-Attention ASR model, introduced in \cite{watanabe2017}, to the multitask learning framework, by attaching a second decoder for subtitle generation and sharing the encoder.

\begin{figure*}[htb]

\begin{minipage}[b]{0.33\linewidth}
  \centering
  \centerline{\includegraphics[height=4.5cm]{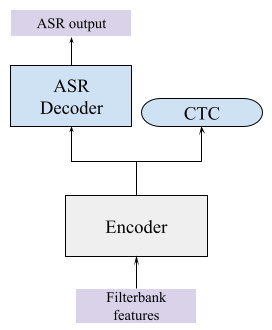}}
  \centerline{(a) E2E ASR model}
  \centerline{    }
  \medskip
\end{minipage}
\hfill
\begin{minipage}[b]{0.33\linewidth}
  \centering
  \centerline{\includegraphics[height=4.5cm]{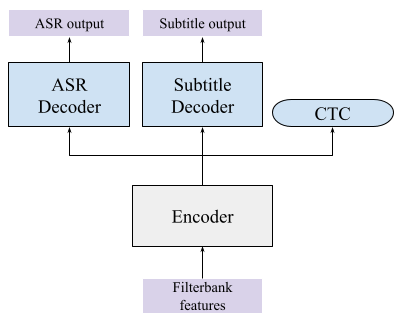}}
  \centerline{(b) Multitask model - ASR + Subtitles}
  \centerline{Independent decoders}
  \medskip
\end{minipage}
\hfill
\begin{minipage}[b]{0.33\linewidth}
  \centering
  \centerline{\includegraphics[height=4.5cm]{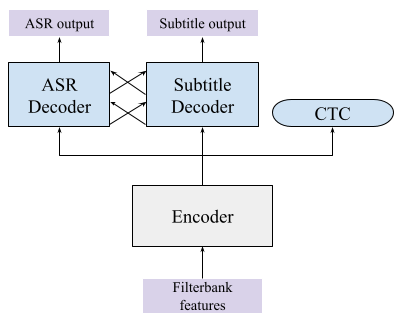}}
  \centerline{(c) Multitask model - ASR + Subtitles}
  \centerline{Cross-connected decoders}
  \medskip
\end{minipage}
\caption{\textit{Model architecture: comparison of (a) a general end-to-end CTC/Attention hybrid ASR model, and (b) the proposed multitask model with two decoders. In (c) the proposed multitask model has additional cross-connections between both decoders.}}
\label{fig:model}
\end{figure*}

\subsection{End-to-end ASR}
End-to-end (E2E) models solve the ASR problem as a sequence-to-sequence problem. First of all, an encoder computes feature representations from the input sequence. The feature extractor in this case is a Transformer \cite{vaswani}, relying on self-attention \cite{SAT-relative} to compute feature representations.

In general, end-to-end architectures either use Connectionist Temporal Classification (CTC) \cite{ctc_graves} to generate an output sequence from the encoder states, or an attention-based mechanism that can attend to different parts of the encoder output to align the input features with the output sequence. Both methods have their strengths and drawbacks. CTC, although efficient, relies on independent Markov assumptions and can only predict monotonic alignments. Pure attention-based methods are hard to train and optimise as there is no constraint on possible alignments.

In this work, we adopt the hybrid CTC/attention architecture \cite{watanabe2017}, that combines both methods, in the baseline ASR system. Figure \ref{fig:model}(a) shows the hybrid model. CTC regularises the encoder by constraining a monotonic alignment and predicting an aligned sequence of output tokens including blanks from the encoder outputs. The encoder outputs could resemble small acoustic units because of the regularisation. The attention-based decoder \cite{vaswani} learns the alignment between the encoder states and the elements of the output sequence. It can attend to the entire encoder output and during decoding the predicted tokens are generated autoregressively, i.e. one-by-one. 

The training objective is a weighted sum of the label smoothed cross-entropy classification loss of the tokens generated by the attention decoder and the CTC objective, respectively denoted by $\mathcal{L}_{att,asr}$ and $\mathcal{L}_{ctc}$.

\begin{equation}
    \mathcal{L}_{asr} = (1-\lambda_{ctc}) \mathcal{L}_{att,asr} + \lambda_{ctc} \mathcal{L}_{ctc}
    \label{eq:Lasr}
\end{equation}

During decoding, the CTC and attention scores are combined in a joint beam search where the produced hypotheses are pre-scored by CTC. 

\subsection{Dual decoder - MTL}
The proposed model is depicted in Figure \ref{fig:model}(b). An additional attention-based decoder branch is added to the encoder. In this multitask model, the encoder is shared but the decoders are completely independent and they both attend to the encoder's states (as the multitask model in \cite{tied2018}). When an auxiliary task is imposed to a model, the performance on the original task might increase if the additional task is useful. The inclusion of a second independent decoder could benefit the learned representation in the shared encoder.

The first decoder, the ASR Decoder, is the same as in the previous section and generates a verbatim transcription. The second decoder, the Subtitle Decoder, generates a subtitle for the input features. 
Due to summarisation and compression, subtitles align differently than verbatim transcriptions. Hence we use an attention-based loss without CTC for the subtitles.
The final training objective $\mathcal{L}_{tot}$ is now a weighted sum of both decoders' attention losses and the CTC objective, with $\mathcal{L}_{asr}$ as defined in Equation \ref{eq:Lasr}.

\begin{equation}
    \mathcal{L}_{tot} = \lambda_{asr}\mathcal{L}_{asr} + \lambda_{subs} \mathcal{L}_{att,subs}
    \label{eq:Ltot}
\end{equation}

During training, batches are filled with utterances from the ASR dataset (i.e. verbatim targets) and utterances from the subtitle dataset, equally mixed per batch. The ASR decoder and CTC losses are backpropagated for the verbatim utterances and masked for the subtitled utterances. 
Likewise, the subtitle decoder loss is backpropagated for the subtitled data and masked for the ASR data.

\subsection{Connected decoders}
\label{sec:cross}
As ASR and subtitling are related tasks, it would seem useful to have some kind of connection between both decoders. Knowing what was said is useful to generate a subtitle, and vice versa. Chaining or cascading the decoders one after another, e.g. an ASR decoder followed by a subtitle decoder, is difficult with autoregressive decoders. Therefore, we investigate the Cross Dual-decoder Transformer proposed in \cite{Le2020}. In this model, there are additional cross-attention blocks added between the decoders, where the decoders can also attend to the previous outputs of the other decoder. Both autoregressive decoders process the data simultaneously and work in a synchronous way. The output of the cross-attention with the other decoder is summed together with or concatenated to the outputs of the decoder's own attention operations \cite{Le2020}. 

The model with cross-connected decoders, depicted in Figure \ref{fig:model}(c), requires parallel data for both the ASR and the subtitle decoder to learn useful cross-attention weights. Therefore, we first generate pseudo-labels with a pre-trained multitask model with independent decoders, i.e. verbatim labels for the subtitle data and subtitles for the ASR data. Second, we finetune the model with additional cross-connections between the decoders on the parallel (pseudo-)labeled data. This is an extension of iterative self-training \cite{selftraining} with pseudo-labels.

During decoding, the ASR and subtitle hypotheses are expanded together as tuples of (ASR, subtitle) combinations in one joint beam. Only valid combinations for the next tokens are considered. The scores per tuple are computed as a weighted combination of the ASR and subtitle hypotheses' scores during the beam search. Equal weights for the ASR and subtitles appear to give the best results.

\section{Experimental setup}
\subsection{Data}
The multitask model is trained on both verbatim ASR data, from an official database, and subtitled data gathered in-house.

\textbf{Verbatim.} The Corpus Gesproken Nederlands (CGN) \cite{CGN_Oostdijk}, or Spoken Dutch Corpus, is a speech database containing 270 hours of Belgian Dutch (Flemish) speech. The complete dataset includes read speech, news reports, lectures, commentaries, and also narrowband telephone speech (8~kHz resampled to 16~kHz) and spontaneous conversational speech (\textit{VL-train-all} in \cite{poncelet}). For evaluation, we use the \textit{dev-other} set of 15h of speech, named \textit{cgn-dev}, which does not contain telephone and conversational speech, and has no overlap in speakers with the training set.

\textbf{Subtitled.} The subtitled dataset contains 720 hours of speech, segmented based on the subtitle timings on screen. The data consists of 190 hours of broadcast TV news, 230 hours of soap TV series, and the remaining 300 hours of speech originate from various talk shows about news, politics, entertainment, etc. For evaluation, we manually annotated a set of 6 hours of speech (around 6500 utterances), named \textit{subs-annot}, from a mixture of show types, and keep it held out of the training set. Finally, for one experiment we have extended the subtitled dataset to 1700 hours of speech gathered from the same and similar sources.

\subsection{Model configuration}
The models are implemented with the ESPnet2 library \cite{watanabe2018espnet} and follow the standard encoder-decoder architecture of end-to-end ASR models. The encoder is a 12-layered Transformer \cite{vaswani} with a Conv2D input layer to transform the input features, and an output dimension of 256. The ASR decoder and the subtitle decoder are both 6-layered Transformer decoders. Every Transformer block has 4 attention heads and 2048 linear units. 

The input features are 80-dimensional Mel filterbank features, concatenated with 3-dimensional pitch features, extracted from windows of 25ms and with a frame shift of 10ms. Utterance-level mean-variance normalisation and SpecAugment \cite{specaug} augmentations are applied to the inputs. The targets are 5000 unigram BPE's, and the BPE (Byte-Pair Encoding) model is trained on both the ASR transcriptions and the subtitles. No language model is applied, not during and not after decoding.


\subsection{Training details}

All models are trained for 100 epochs with a batch size of 32 and gradient accumulation over 8 steps. We use the Adam optimizer \cite{adam} with a learning rate schedule for Transformers \cite{vaswani} with 25000 warm-up steps and a peak learning rate of 0.004 for models trained from scratch and 0.0004 for ASR-initialised models. For evaluation, we average the 10 models with the best ASR accuracy on the validation set. The CTC weight $\lambda_{ctc}$ is set to 0.3, both during training and decoding. The cross-entropy label smoothing weight is set to 0.1. Decoding is implemented as beam search with beam size 20. 

\section{Results}

In this section we will describe the experiments conducted to evaluate the proposed model and discuss the results. We examine the performance of the multitask model on a speech recognition task, and compare it to the ASR model without subtitle decoder. The models are evaluated based on the Word Error Rate (WER) on a regular ASR test set \textit{cgn-dev} and on the spontaneous test set \textit{subs-annot}. Furthermore, the capabilities of automatic subtitling of the multitask model are explored by comparing the output hypothesis of the subtitle decoder to the real subtitle on screen. As metric, we utilise the smoothed BLEU-4 score \cite{BLEU} with uniform weights. Finally, we investigate if finetuning with generated pseudo-labels can yield further improvements. We compare the multitask model with independent decoders to the multitask model with cross-attention connected decoders on ASR and the subtitling task.

\subsection{Verbatim Speech Recognition}

\subsubsection{Multitask model from scratch}
First of all, we examine the effect of adding the additional subtitle decoder to the ASR model. The importance of both decoders can be weighted in the loss function (Equation \ref{eq:Ltot}) with the hyperparameters $\lambda_{subs}$ and $\lambda_{asr}$. Table \ref{tab:res1} shows the results for different combinations of both weights. The first line, i.e. the model with $\lambda_{subs}=0$, is the baseline ASR model. 

It is important to note that training on the joint set of ASR and subtitle data, as if it were all verbatim data, always diverges, irrespective of the learning rate. We postulate that the inexact transcripts could impair the CTC alignment and the classification, and the inaccurate timings are difficult to incorporate in verbatim speech recognition. This is the key advantage of the proposed multitask model.

Table \ref{tab:res1} shows that the additional decoder can slightly improve the WER on the regular \textit{cgn-dev} and strongly improve the WER on the spontaneous \textit{subs-annot}, because the encoder is improved by incorporating the subtitle data for the subtitling task. Using equal weights in the loss function seems most beneficial in this setup. If the weight of the subtitle decoder $\lambda_{subs}$ is too large, the model tends to diverge. It is a hyperparameter that needs some tuning, but always improves results when chosen between a wide range.

\begin{table}[ht]
    \centering
    \begin{tabular}{c|c||c|c}
        \toprule[1.5pt]
        \multicolumn{2}{c||}{\textbf{Decoder}} & \multicolumn{2}{c}{\textbf{WER (\%)}}\\
        $\mathbf{\lambda_{asr}}$ & $\mathbf{\lambda_{subs}}$ & \textit{cgn-dev} & \textit{subs-annot} \\
        \midrule[1pt]
        1.0 & 0.0 & 15.90 & 37.61 \\
        \midrule[0.1pt]
        0.9 & 0.1 & 16.54 & 36.76 \\
        0.7 & 0.3 & 14.50 & 32.47 \\
        0.5 & 0.5 & \textbf{14.29} & \textbf{30.48} \\
        0.3 & 0.7 & 79.98 (div) & 83.18 (div) \\
        0.1 & 0.9 & 103.57 (div) & 109.56 (div) \\
        \bottomrule[1.5pt]
        
    \end{tabular}
    \caption{\textit{WER for multitask models trained from scratch with varying weights in the objective function. The first row shows the single decoder ASR model as baseline. Diverged models are annotated with (div).}}
    \label{tab:res1}
\end{table}

To get a sense of the difference between the ASR and subtitle decoder output, Table \ref{tab:str} shows a sample utterance transcription from the spontaneous test set and the generated hypotheses from both decoders, with the model in Table \ref{tab:res1} using $\lambda_{subs}=0.5$. Although for this utterance the reference subtitle is much shorter and extremely hard to predict, it is clear that the subtitle decoder output is cleaner, e.g. no apostrophies for abbreviating words and no hesitations (`uh'). We found this to be a recurring theme during analysis of the subtitle output, with additional normalisation of dialectal words (e.g. `gij' becomes `jij') and the decoder leaving out repetitions due to stammering.

\begin{table*}[ht]
    \centering
    \begin{tabular}{c c}
        \toprule[1.5pt]
        \textbf{REF} & en 't is net op dat moment dat het virus eigenlijk toeslaat da's een uh eerste feit \\
        (verbatim) & \textit{and it's right at that moment that the virus actually hits that's an uh first fact} \\
        \midrule[0.1pt]
        \textbf{REF} & dan slaat het virus toe dat is een eerste feit \\
        (subtitle) & \textit{then the virus hits that is a first fact} \\
        \midrule[1pt]
        \textbf{ASR} & 't is net op dat moment dat het virus eigenlijk toelaat was een uh feit \\
         hypothesis & \textit{it's right at that moment that the virus actually permits was an uh fact}\\
         \midrule[0.1pt]
         \textbf{SUB} & het is net op dat moment dat het virus eigenlijk toelaat als een feit \\
         hypothesis & \textit{it is right at that moment that the virus actually permits as a fact}\\
         \bottomrule[1.5pt]
    \end{tabular}
    \caption{\textit{Comparison of the reference ASR transcription and the reference subtitle transcription to the hypotheses generated by the outputs of the ASR decoder (ASR) and the subtitle decoder (SUB). The utterance is part of the \textit{subs-annot} test set. Translation in English is added for the reader.}}
    \label{tab:str}
\end{table*}

\subsubsection{Adapting pre-trained ASR models to multitask models}

Second, we examine if a pre-trained single-decoder ASR model can be improved with the additional decoder, or if it would harm the result on the ASR task. The ASR model is a state-of-the-art E2E ASR model \cite{watanabe2017}. Table \ref{tab:res2} shows the results when the encoder is initialised with a pre-trained ASR encoder, and the ASR and/or the subtitle decoder branch are initialised with the parameters from the decoder of a pre-trained ASR model. The baseline model of the previous section is used as pre-trained ASR model to initialise the weights.


The WER on both sets strongly improves by extending a pre-trained ASR model with the additional subtitle decoder and using the subtitle data. Both the ASR decoder and subtitle decoder branch can be initialised with an ASR model. 

\begin{table}[ht]
    \centering
    \begin{tabular}{c|c|c|c||c|c}
        \toprule[1.5pt]
        \multicolumn{2}{c|}{\textbf{Decoder}} & \multicolumn{2}{c||}{\textbf{Initialisation}} &  \multicolumn{2}{c}{\textbf{WER (\%)}} \\
        $\mathbf{\lambda_{asr}}$ & $\mathbf{\lambda_{subs}}$ & ASR & Subs & \textit{cgn-dev} & \textit{subs-annot} \\
        \midrule[1pt]
        1.0 & 0.0 & / & / & 15.90 & 37.61 \\
        \midrule[0.1pt]
        0.5 & 0.5 & / & / & 14.29 & 30.48 \\
        0.5 & 0.5 & ASR & / & 12.27 & \textbf{28.33} \\
        0.5 & 0.5 & ASR & ASR & \textbf{12.22} & 28.38 \\
        \bottomrule[1.5pt]
    \end{tabular}
    \caption{\textit{WER for multitask models with initialisation of the ASR decoder (ASR) and/or subtitle decoder (Subs) with the weights of a pre-trained ASR decoder. The models in the first two rows are not initialised and trained from scratch. The models in the bottom two rows both use a pre-trained ASR encoder. The subtitle dataset contains 700h of data.}}
    \label{tab:res2}
\end{table}

\subsubsection{Using more subtitled data}
Finally, we examine if using more subtitled data would additionally improve the results. We compare the ASR performance of a model using 700 hours of subtitled data in Table \ref{tab:res2} to a model using 1700 hours of subtitled data in Table \ref{tab:res3}. The verbatim ASR dataset remains the same.

The results on 1700 hours (using the same settings as in the previous experiments) are comparable to the results on 700 hours of subtitled data, but not better. We notice the same trend of improvements across the board for the different models.
The first line of the table suggests that further finetuning of $\lambda_{asr}$ might yield further improvements.

\begin{table}[ht]
    \centering
    \begin{tabular}{c|c|c|c||c|c}
        \toprule[1.5pt]
        \multicolumn{2}{c|}{\textbf{Decoder}} & \multicolumn{2}{c||}{\textbf{Initialisation}} &  \multicolumn{2}{c}{\textbf{WER (\%)}} \\
        $\mathbf{\lambda_{asr}}$ & $\mathbf{\lambda_{subs}}$ & ASR & Subs & \textit{cgn-dev} & \textit{subs-annot} \\
        \midrule[1pt]
        0.7 & 0.3 & / & / & 14.37 & 32.84 \\
        0.5 & 0.5 & / & / & 16.26 & 33.45 \\
        0.5 & 0.5 & ASR & / & 12.56 & 30.04 \\
        0.5 & 0.5 & ASR & ASR & \textbf{12.23} & \textbf{29.86} \\
        \bottomrule[1.5pt]
    \end{tabular}
    \caption{\textit{WER for multitask models trained on the extended 1700h subtitle dataset, with or without pre-trained initialisation.}}
    \label{tab:res3}
\end{table}

\subsection{Subtitling}
We evaluate the output of the subtitle decoder with a BLEU score against the reference on-screen subtitle in the test set. Table \ref{tab:res4} summarises the results of all previous models. For the baseline (i.e. the first row), we use the output of the ASR decoder. This gives a notice of how close the subtitles are to verbatim transcriptions.

For almost all cases, the improvements in BLEU score follow the improvements in WER as in the previous section. The double-initialised dual-decoder model achieves the highest BLEU scores. Additionally, we notice that for most cases there is no improvement with the model using 1700 hours of subtitle data over the model using 700 hours of subtitle data. This might suggest that the additional 1000 hours of data, mostly coming from different sources, is slightly mismatched with the test set or contains more non-Dutch transcriptions.

\begin{table}[ht]
    \centering
    \begin{tabular}{c|c|c|c|c||c}
        \toprule[1.5pt]
        \textbf{\# Hours} & \multicolumn{2}{c|}{\textbf{Decoder}} & \multicolumn{2}{c||}{\textbf{Initialisation}} & \textbf{BLEU (\%)} \\
        \textit{Subs} & $\mathbf{\lambda_{asr}}$ & $\mathbf{\lambda_{subs}}$ & ASR & Subs & \textit{subs-annot} \\
        \midrule[1pt]
        0 & 1.0 & 0.0 & / & / & 30.84 \\
        \midrule[0.1pt]
        \multirow{7}{*}{700} & 0.9 & 0.1 & / & / & 40.86 \\
        & 0.7 & 0.3 & / & / & 43.63 \\
        & 0.5 & 0.5 & / & / & 45.65 \\
        & 0.3 & 0.7 & / & / & 15.50 (div) \\
        & 0.1 & 0.9 & / & / & 8.08 (div) \\
        & 0.5 & 0.5 & ASR & / & 36.51 \\
        & 0.5 & 0.5 & ASR & ASR & \textbf{46.81} \\
        \midrule[0.1pt]
        \multirow{4}{*}{1700} & 0.7 & 0.3 & / & / & 41.66 \\
        & 0.5 & 0.5 & / & / & 41.41 \\
        & 0.5 & 0.5 & ASR & / & 43.60 \\
        & 0.5 & 0.5 & ASR & ASR & \textbf{43.85} \\
        \bottomrule[1.5pt]
    \end{tabular}
    \caption{\textit{BLEU scores for multitask models with varying weights in the objective function and possible initialisation. The first row shows the single-decoder ASR model. Diverged models are annotated with (div).}}
    \label{tab:res4}
\end{table}

\subsection{Connected decoders}
Section \ref{sec:cross} showed a method to connect both decoders of the multitask model. Because the connected decoders require parallel data, we generate pseudo-labels with the ASR decoder for the subtitled data and with the subtitle decoder for the verbatim data. For pseudo-labeling we decode with the best-performing model, i.e. the model of the last row in Table \ref{tab:res2}, which is also used for initialisation of all weights except for the weights in the cross-attention part between the decoders. The loss weights $\lambda_{asr}$ and $\lambda_{subs}$ are both kept as 0.5. We compare the multitask model with cross-connected decoders to the multitask model with independent decoders (as in previous sections), starting from the same initialisation and further trained with pseudo-labels. Table \ref{tab:res5} shows the ASR results and Table \ref{tab:res6} shows the subtitling results.

Further finetuning of the multitask model with independent decoders on the generated pseudo-labels (row 2 in Table \ref{tab:res5}), starting from the pre-trained multitask model (row 1 in Table \ref{tab:res5}), slightly improves the WER on both sets, at the expense of a drop in BLEU score.

Finetuning the pre-trained multitask model (row 1 in Table \ref{tab:res5}) with additional cross-attention connections between the decoders is not able to improve the WER on the regular \textit{cgn-dev} test set. 
This might be because the ASR decoder cross-attention and the Subtitle-to-ASR cross-attention are merged (either summed up or concatenated and linearly projected), which could implicitly give an extra weight to the subtitles in the ASR decoder branch, which might have a slight negative impact on the ASR decoder in very clean settings due to the domain mismatch.
However, when the encoder is kept frozen during finetuning and only the decoders are finetuned, we do obtain the best ASR performance on the spontaneous \textit{subs-annot} test set and by far the best BLEU score. 

\begin{table}[ht]
    \centering
    \begin{tabular}{c|c|c|c||c|c}
        \toprule[1.5pt]
        \multicolumn{3}{c|}{\textbf{Model}} & \textbf{PL} & \multicolumn{2}{c}{\textbf{WER (\%)}} \\
        Cross & Method & Freeze & & \textit{cgn-dev} & \textit{subs-annot} \\
        \midrule[1pt]
        No & / & / & No & 12.22 & 28.38 \\
        No & / & / & Yes & \textbf{12.04} & 28.33 \\
        \midrule[0.1pt]
        Yes & Sum & / & Yes & 12.32 & 28.44 \\
        Yes & Concat & / & Yes & 12.37 & 28.39 \\
        Yes & Concat & Enc. & Yes & 13.46 & \textbf{27.69} \\
        
        \bottomrule[1.5pt]
    \end{tabular}
    \caption{\textit{WER for multitask models with independent decoders and with cross-connected decoders. The decoder attention outputs and the inter-decoder cross-attention outputs can be concatenated or summed. The models can be further trained with pseudo-labels (PL) generated by the decoders for missing targets. The encoder can be kept frozen during finetuning with the pseudo-labels.}}
    \label{tab:res5}
\end{table}

\begin{table}[ht]
    \centering
    \begin{tabular}{c|c|c|c||c}
        \toprule[1.5pt]
        \multicolumn{3}{c|}{\textbf{Model}} & \textbf{PL} &  \textbf{BLEU} \\
        Cross & Method & Freeze & & \textit{subs-annot} \\
        \midrule[1pt]
        No & / & / & No & 46.81 \\
        No & / & / & Yes & 36.44 \\
        \midrule[0.1pt]
        Yes & Sum & / & Yes & 45.97 \\
        Yes & Concat & / & Yes & 45.38 \\
        Yes & Concat & Encoder & Yes & \textbf{51.11} \\
        \bottomrule[1.5pt]
    \end{tabular}
    \caption{\textit{BLEU scores for multitask models with independent decoders and with cross-connected decoders. The decoder attention outputs and the inter-decoder cross-attention outputs can be concatenated or summed. The models can be further trained with pseudo-labels (PL) generated by the decoders for missing targets. The encoder can be kept frozen during finetuning with the pseudo-labels.}}
    \label{tab:res6}
\end{table}

\section{Conclusion}
We have proposed a model for joint automatic speech recognition and subtitle generation, that is able to leverage subtitled data to improve ASR performance on both regular and spontaneous speech through multitask learning of both tasks. A pre-trained ASR model can easily be adapted with a second decoder, and its performance can be improved even with the inaccurate subtitle transcriptions. 

In this work, we use a dual-decoder model with independent autoregressive decoders and a dual-decoder model with synchronously cross-connected autoregressive decoders.

In future work, we will investigate non-autoregressive decoders 
to chain the decoders in a cascaded way. We will also look at the MGB challenge for evaluation of the models. The multitask models might also be improved through better alignment of the subtitles' endpoints with the input speech.

\section{Acknowledgement}

This research received funding from the Flemish Government under the ``Onderzoeksprogramma Artificiële Intelligentie (AI) Vlaanderen" programme and was supported by Research Foundation Flanders (FWO) under grant S004923N of the SBO programme.


\bibliographystyle{IEEEbib}
\bibliography{main}

\begin{thebibliography}{10}

\bibitem{librispeech}
V.~{Panayotov}, G.~{Chen}, D.~{Povey}, and S.~{Khudanpur},
\newblock ``{Librispeech}: An {A}{S}{R} corpus based on public domain audio
  books,''
\newblock in {\em 2015 IEEE International Conference on Acoustics, Speech and
  Signal Processing (ICASSP)}, 2015, pp. 5206--5210.

\bibitem{wav2vec2}
Alexei Baevski, Yuhao Zhou, Abdelrahman Mohamed, and Michael Auli,
\newblock ``wav2vec 2.0: A framework for self-supervised learning of speech
  representations,''
\newblock in {\em Advances in Neural Information Processing Systems (NIPS)},
  2020, vol.~33, pp. 12449--12460.

\bibitem{hubert}
Wei-Ning Hsu, Benjamin Bolte, Yao-Hung~Hubert Tsai, Kushal Lakhotia, Ruslan
  Salakhutdinov, and Abdelrahman Mohamed,
\newblock ``Hubert: Self-supervised speech representation learning by masked
  prediction of hidden units,''
\newblock {\em IEEE/ACM Transactions on Audio, Speech, and Language Processing
  (TASLP)}, vol. 29, pp. 3451--3460, 2021.

\bibitem{wavlm}
Sanyuan Chen, Chengyi Wang, Zhengyang Chen, Yu~Wu, Shujie Liu, Zhuo Chen, Jinyu
  Li, Naoyuki Kanda, Takuya Yoshioka, Xiong Xiao, Jian Wu, Long Zhou, Shuo Ren,
  Yanmin Qian, Yao Qian, Jian Wu, Michael Zeng, Xiangzhan Yu, and Furu Wei,
\newblock ``{WavLM}: Large-scale self-supervised pre-training for full stack
  speech processing,''
\newblock {\em IEEE Journal of Selected Topics in Signal Processing (JSTSP)},
  pp. 1--14, 2022.

\bibitem{selftraining}
Qiantong Xu, Alexei Baevski, Tatiana Likhomanenko, Paden Tomasello, Alexis
  Conneau, Ronan Collobert, Gabriel Synnaeve, and Michael Auli,
\newblock ``Self-training and pre-training are complementary for speech
  recognition,''
\newblock in {\em 2021 IEEE International Conference on Acoustics, Speech and
  Signal Processing (ICASSP)}, 2021, pp. 3030--3034.

\bibitem{XLSR}
Arun Babu, Changhan Wang, Andros Tjandra, Kushal Lakhotia, Qiantong Xu, Naman
  Goyal, Kritika Singh, Patrick von Platen, Yatharth Saraf, Juan Pino, Alexei
  Baevski, Alexis Conneau, and Michael Auli,
\newblock ``{XLS-R}: Self-supervised cross-lingual speech representation
  learning at scale,''
\newblock in {\em Interspeech 2022}, 2022.

\bibitem{MTL1}
Rich Caruana,
\newblock ``Multitask learning,''
\newblock {\em Machine learning}, vol. 28, no. 1, pp. 41--75, 1997.

\bibitem{MTL2}
Sebastian Ruder,
\newblock ``An overview of multi-task learning in deep neural networks,''
\newblock {\em CoRR}, vol. abs/1706.05098, 2017.

\bibitem{tied2018}
Antonios Anastasopoulos and David Chiang,
\newblock ``Tied multitask learning for neural speech translation,''
\newblock in {\em Proceedings of the 2018 Conference of the North {A}merican
  Chapter of the Association for Computational Linguistics (NAACL): Human
  Language Technologies}, 2018, vol.~1, pp. 82--91.

\bibitem{Le2020}
Hang Le, Juan~Miguel Pino, Changhan Wang, Jiatao Gu, Didier Schwab, and Laurent
  Besacier,
\newblock ``Dual-decoder transformer for joint automatic speech recognition and
  multilingual speech translation,''
\newblock in {\em Proceedings of the 28th International Conference on
  Computational Linguistics (COLING)}, 2020, pp. 3520--3533.

\bibitem{interactive2020}
Yuchen Liu, Jiajun Zhang, Hao Xiong, Long Zhou, Zhongjun He, Hua Wu, Haifeng
  Wang, and Chengqing Zong,
\newblock ``Synchronous speech recognition and speech-to-text translation with
  interactive decoding,''
\newblock {\em Proceedings of the AAAI Conference on Artificial Intelligence},
  vol. 34, pp. 8417--8424, 2020.

\bibitem{poncelet}
Jakob Poncelet and Hugo Van~hamme,
\newblock ``Comparison of self-supervised speech pre-training methods on
  {F}lemish {D}utch,''
\newblock in {\em 2021 IEEE Automatic Speech Recognition and Understanding
  Workshop (ASRU)}, 2021, pp. 169--176.

\bibitem{Bang2020}
Jeong-Uk Bang, Mu-Yeol Choi, Sang-Hun Kim, and O.~W. Kwon,
\newblock ``Automatic construction of a large-scale speech recognition database
  using multi-genre broadcast data with inaccurate subtitle timestamps,''
\newblock {\em IEICE Transactions on Information and Systems}, vol. 103-D, pp.
  406--415, 2020.

\bibitem{Lamel}
Lori Lamel, Jean-Luc Gauvain, and Gilles Adda,
\newblock ``Lightly supervised acoustic model training,''
\newblock in {\em Proceedings of the ISCA ITRW ASR'2000}, 2000, pp. 150--154.

\bibitem{Bang2017}
Jeong-Uk Bang, Mu-Yeol Choi, Sang-Hun Kim, and Oh-Wook Kwon,
\newblock ``Improving speech recognizers by refining broadcast data with
  inaccurate subtitle timestamps,''
\newblock in {\em Interspeech 2017}, 2017, pp. 2929--2933.

\bibitem{CU_MGB1}
P.~C. Woodland, X.~Liu, Y.~Qian, C.~Zhang, M.~J.~F. Gales, P.~Karanasou,
  P.~Lanchantin, and L.~Wang,
\newblock ``Cambridge {U}niversity transcription systems for the multi-genre
  broadcast challenge,''
\newblock in {\em 2015 IEEE Workshop on Automatic Speech Recognition and
  Understanding (ASRU)}, 2015, pp. 639--646.

\bibitem{Shintaro_ICASSP21}
Shintaro Ando and Hiromasa Fujihara,
\newblock ``Construction of a large-scale {J}apanese {ASR} corpus on {TV}
  recordings,''
\newblock in {\em 2021 IEEE International Conference on Acoustics, Speech and
  Signal Processing (ICASSP)}, 2021, pp. 6948--6952.

\bibitem{LIUM_MGB2}
Natalia Tomashenko, Kévin Vythelingum, Anthony Rousseau, and Yannick Estève,
\newblock ``{LIUM} {ASR} systems for the 2016 multi-genre broadcast {A}rabic
  challenge,''
\newblock in {\em 2016 IEEE Spoken Language Technology Workshop (SLT)}, 2016,
  pp. 285--291.

\bibitem{SU_MGB1}
Oscar Saz, Mortaza Doulaty, Salil Deena, Rosanna Milner, Raymond W.~M. Ng,
  Madina Hasan, Yulan Liu, and Thomas Hain,
\newblock ``The 2015 {S}heffield system for transcription of multi-genre
  broadcast media,''
\newblock in {\em 2015 {IEEE} Workshop on Automatic Speech Recognition and
  Understanding ({ASRU})}, 2015.

\bibitem{iwslt1}
Danni Liu, Jan Niehues, and Gerasimos Spanakis,
\newblock ``Adapting end-to-end speech recognition for readable subtitles,''
\newblock in {\em Proceedings of the 17th International Conference on Spoken
  Language Translation (IWSLT)}, 2020, pp. 247--256.

\bibitem{CRIM_MGB1}
Vishwa Gupta, Paul Deléglise, Gilles Boulianne, Yannick Estève, Sylvain
  Meignier, and Anthony Rousseau,
\newblock ``{CRIM} and {LIUM} approaches for multi-genre broadcast media
  transcription,''
\newblock in {\em 2015 IEEE Workshop on Automatic Speech Recognition and
  Understanding (ASRU)}, 2015, pp. 681--686.

\bibitem{icalt1}
Xiaoyin Che, Sheng Luo, Haojin Yang, and Christoph Meinel,
\newblock ``Automatic lecture subtitle generation and how it helps,''
\newblock in {\em 2017 IEEE 17th International Conference on Advanced Learning
  Technologies (ICALT)}, 2017, pp. 34--38.

\bibitem{MGB1}
P~Bell, M~J~F Gales, T~Hain, J~Kilgour, P~Lanchantin, X~Liu, A~McParland,
  S~Renals, O~Saz, M~Wester, and P~C Woodland,
\newblock ``The {MGB} challenge: Evaluating multi-genre broadcast media
  recognition,''
\newblock in {\em 2015 IEEE Workshop on Automatic Speech Recognition and
  Understanding (ASRU)}, 2015, pp. 687--693.

\bibitem{MGB2}
Ahmed Ali, Peter Bell, James Glass, Yacine Messaoui, Hamdy Mubarak, Steve
  Renals, and Yifan Zhang,
\newblock ``The {MGB}-2 challenge: Arabic multi-dialect broadcast media
  recognition,''
\newblock in {\em 2016 IEEE Spoken Language Technology Workshop (SLT)}, 2016,
  pp. 279--284.

\bibitem{IberSpeechRTVE}
Eduardo Lleida, Alfonso Ortega, Antonio Miguel, Virginia Bazán-Gil, Carmen
  Pérez, Manuel Gómez, and Alberto de~Prada,
\newblock ``Albayzin 2018 evaluation: The {IberSpeech-RTVE} challenge on speech
  technologies for {S}panish broadcast media,''
\newblock {\em Applied Sciences}, vol. 9, no. 24, 2019.

\bibitem{WeaklyICASSP2020}
Kritika Singh, Dmytro Okhonko, Jun Liu, Yongqiang Wang, Frank Zhang, Ross
  Girshick, Sergey Edunov, Fuchun Peng, Yatharth Saraf, Geoffrey Zweig, and
  Abdelrahman Mohamed,
\newblock ``Training {ASR} models by generation of contextual information,''
\newblock in {\em 2020 IEEE International Conference on Acoustics, Speech and
  Signal Processing (ICASSP)}, 2020, pp. 7864--7868.

\bibitem{lightsup_align}
Oscar Saz, Salil Deena, Mortaza Doulaty, Madina Hasan, Bilal Khaliq, Rosanna
  Milner, Raymond W.~M. Ng, Julia Olcoz, and Thomas Hain,
\newblock ``Lightly supervised alignment of subtitles on multi-genre
  broadcasts,''
\newblock {\em Multimedia Tools and Applications}, vol. 77, 2018.

\bibitem{watanabe2017}
Shinji Watanabe, Takaaki Hori, Suyoun Kim, John~R. Hershey, and Tomoki Hayashi,
\newblock ``Hybrid {CTC}/{A}ttention architecture for end-to-end speech
  recognition,''
\newblock {\em IEEE Journal of Selected Topics in Signal Processing (JSTSP)},
  vol. 11, no. 8, pp. 1240--1253, 2017.

\bibitem{vaswani}
Ashish Vaswani, Noam Shazeer, Niki Parmar, Jakob Uszkoreit, Llion Jones,
  Aidan~N Gomez, \L~ukasz Kaiser, and Illia Polosukhin,
\newblock ``Attention is all you need,''
\newblock in {\em Advances in Neural Information Processing Systems (NIPS)},
  2017, vol.~30.

\bibitem{SAT-relative}
Peter Shaw, Jakob Uszkoreit, and Ashish Vaswani,
\newblock ``Self-attention with relative position representations,''
\newblock in {\em Proceedings of the 2018 Conference of the North {A}merican
  Chapter of the Association for Computational Linguistics (NAACL): Human
  Language Technologies}, 2018, vol.~2, pp. 464--468.

\bibitem{ctc_graves}
Alex Graves, Santiago Fernández, and Faustino Gomez,
\newblock ``Connectionist temporal classification: Labelling unsegmented
  sequence data with recurrent neural networks,''
\newblock in {\em Proceedings of the International Conference on Machine
  Learning (ICML)}, 2006, pp. 369--376.

\bibitem{CGN_Oostdijk}
Nelleke Oostdijk,
\newblock ``The {S}poken {D}utch {C}orpus: Overview and first evaluation,''
\newblock in {\em Proceedings of the International Conference on Language
  Resources and Evaluation (LREC)}, 2000, vol.~2.

\bibitem{watanabe2018espnet}
Shinji Watanabe, Takaaki Hori, Shigeki Karita, Tomoki Hayashi, Jiro Nishitoba,
  Yuya Unno, Nelson {Enrique Yalta Soplin}, Jahn Heymann, Matthew Wiesner,
  Nanxin Chen, Adithya Renduchintala, and Tsubasa Ochiai,
\newblock ``{ESPnet}: End-to-end speech processing toolkit,''
\newblock in {\em Interspeech 2018}, 2018, pp. 2207--2211.

\bibitem{specaug}
Daniel~S. Park, William Chan, Yu~Zhang, Chung-Cheng Chiu, Barret Zoph, Ekin~D.
  Cubuk, and Quoc~V. Le,
\newblock ``{SpecAugment}: A simple data augmentation method for automatic
  speech recognition,''
\newblock in {\em Interspeech 2019}, 2019.

\bibitem{adam}
Diederik~P. Kingma and Jimmy Ba,
\newblock ``Adam: {A} method for stochastic optimization,''
\newblock in {\em 2015 3rd International Conference on Learning Representations
  (ICLR)}, 2015.

\bibitem{BLEU}
Kishore Papineni, Salim Roukos, Todd Ward, and Wei-Jing Zhu,
\newblock ``{BLEU}: a method for automatic evaluation of machine translation,''
\newblock in {\em Proceedings of the 40th Annual Meeting of the Association for
  Computational Linguistics}, 2002, pp. 311--318.

\end{thebibliography}

\end{document}